\documentclass{elsart}

\usepackage{graphicx}
\usepackage{amsmath}
\usepackage{amsfonts}
\usepackage{amssymb}
\usepackage{epsfig}
\usepackage{verbatim} 
\usepackage{color} 

\newcommand{\feyndag}[1]{\ensuremath{\not{\hspace{-0.75mm} #1}}}

\begin{document}

\begin{frontmatter}

\title{Support of generalized parton distributions in Bethe-Salpeter models of hadrons}

\author{A. Van Dyck\ead{Annelies.VanDyck@UGent.be}},
\author{T. Van Cauteren\thanksref{leaveUCM}},
\author{J. Ryckebusch}

\address{Department of Subatomic and
  Radiation Physics, Ghent University, Proeftuinstraat 86, B-9000
  Gent, Belgium}

\thanks[leaveUCM]{Current address: Universidad Complutense de Madrid,
  Ciudad Universitaria s/n, E-28040 Madrid}

\begin{abstract}
The proper support of generalized parton distributions from relativistic
constituent quark models with pointlike constituents is
studied. The correct support is guaranteed when the vertex function
does not depend on the relative minus-momentum. We show that including quark
interactions in models with pointlike constituent quarks might
lead to a support problem. A computation of the magnitude of the
support problem in the Bonn relativistic constituent quark model is presented.
\end{abstract}

\end{frontmatter}

Historically, form factors and parton distribution functions have
played a key role in hadron physics. In 1997, a
new research tool emerged in the framework of the generalized parton
distributions (GPDs) \cite{ji:first}, providing us with a natural
unification of form factors and parton distribution functions
\cite{belitskymueller:ff}. GPDs describe the soft, \textit{i.e.}
non-perturbative hadronic, scattering part of Hard Exclusive Meson
Production (HEMP) and Deeply Virtual Compton Scattering (DVCS). Today,
a lot of experimental and theoretical effort is devoted to the study of these
generalized parton distributions, of which is hoped that they will
shed light on the origin of the spin of hadrons in general, and the
nucleon in particular. 

\begin{figure}
\centering
\includegraphics[width=0.7\textwidth]{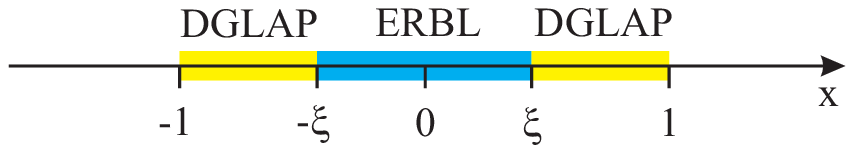}
\caption{Pictorial indication of the supported regions.}
\label{Fig:regions}
\end{figure}

At present, calculating GPDs from first QCD principles is not feasible,
and one relies on chiral perturbation theory \cite{diehl:chpt},
lattice QCD \cite{lattice:GPD}, parametrizations
\cite{Guidal:2004nd} or phenomenological models 
\cite{ji:mitbaggpd,boffi:linking,scopetta&vento:gpdincqm,scopetta:comp,noguera:bsapp,luik:gpd,choi1,choi2,tibmil1,tibmil2,tibmil3}. 
This letter focuses on the latter class. 
In order to compare the GPDs computed in a phenomenological model with
the DVCS and HEMP data, a $Q^2$-evolution needs
to be performed \cite{jaffe:q2}. The evolution equations depend highly
on the kinematic region of the process. More specifically, one can 
distinguish between the DGLAP and ERBL regions, where the
$Q^2$-evolution is respectively governed by the DGLAP and the ERBL
equations. As indicated schematically in Fig. \ref{Fig:regions},
parton model constraints make GPDs vanish outside these two regions
\cite{diehl:review,jaffe:pdftwist4}. A model is said to have the
correct support, when it can resolve the DGLAP and ERBL regions, and
has a vanishing GPD outside this region. 
Developing such a model 
is far from straightforward. In this letter we 
will show how the support can arise naturally in quark models that
are based on the Bethe-Salpeter approach. Including quark dynamics in
the model, however, is likely to destroy the correct support.  

In the following, we will focus on the GPD of a pseudoscalar meson
such as the pion, considering the case where the partons do not
transfer helicity. Pseudoscalar mesons have only one such generalized  
quark distribution associated with them, whereas the nucleon has
four. The theoretical analysis of the support region is, therefore,
less cumbersome for the pion GPD than for the nucleon ones. 

GPDs are non-diagonal matrix elements of a bilocal field operator on
the light cone. For partons with spin $1/2$ (quarks) in a pseudoscalar
meson, the GPD associated with helicity conserving partons is defined
as follows \cite{diehl:review}:  
\begin{equation}
  H_{\pi}(x,\xi,t) =
  \frac{1}{2}\int\frac{\mathrm{d}z^-}{2\pi}e^{ix\tilde{P}^+z^-}\left.\langle\bar{P}^{\prime}
  |\bar{\psi}(-\frac{z}{2})\gamma^{+}\psi(\frac{z}{2})|\bar{P}\rangle\right|_{z^+=0,
  \mathbf{z_{\bot}}=\mathbf{0}}\,.
  \label{eq:defgpd}
\end{equation} 
In this equation, $\tilde{P} =
\frac{\bar{P}^{\prime}+\bar{P}}{2}$. Definition (\ref{eq:defgpd}) uses
light-cone coordinates. A fourmomentum $p=(p^0,p^1,p^2,p^3)$ can be
written in light-cone coordinates as $p = [p^+,p^-,\boldsymbol{p}_{\perp}]$
with $p^\pm = (p^0 \pm p^3)/\sqrt{2}$ and $\boldsymbol{p}_{\perp} = (p^1,p^2)$. 

\begin{figure}
\centering
\includegraphics[width=0.7\textwidth]{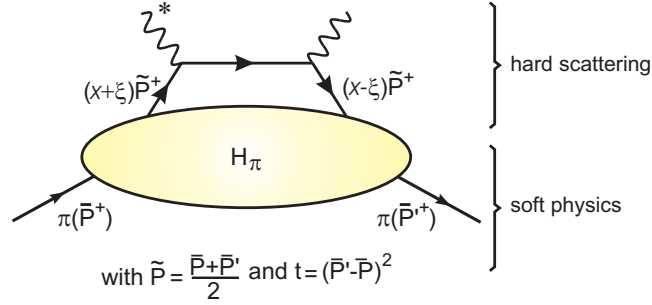}
\caption{Feynman diagram of DVCS, with definition of the kinematical
  variables $x$, $\xi$ and $t$.}
\label{Fig:kinematics}
\end{figure}

In order to calculate the GPD, several choices regarding the kinematics of
the problem have to be made. In fact, two choices are used in
literature \cite{ji:vars,radyushkin:vars}. We make use of the
definitions from \cite{ji:vars}, illustrated in Fig. \ref{Fig:kinematics}
for the DVCS process: $x$ denotes the fraction of the average pion
plus-momentum that is reabsorbed by the meson, while the skewedness
$\xi$ is a  measure for the plus-momentum that is lost in the process:
\begin{equation}
  \xi = \frac{\bar{P}^+ - \bar{P}^{\prime+}}{\bar{P}^+ + \bar{P}^{\prime
      +}}\,.
\end{equation}
$\xi$ is defined on the interval $[0,\xi_{max}]$ with
\begin{equation}
  \xi_{max} = \sqrt{\frac{-t}{4 M_{\pi}^2 - t}} \; ,
\end{equation}
where $t = \Delta^2 = (\bar{P}^{\prime} - \bar{P})^2$. The
abovementioned DGLAP and ERBL regions are defined as follows: 
\begin{itemize}
\item $x \in [-1,-\xi]$: DGLAP region (emission and absorption of
  antiquark);
\item $x \in [-\xi,\xi]$: ERBL region (emission of both quark and
  antiquark);
\item $x \in [\xi,1]$: DGLAP region (emission and absorption of
  quark).
\end{itemize}
This is depicted schematically in Fig. \ref{Fig:regions}. As mentioned
before, the GPD should vanish outside these regions, \textit{i.e.} for
$|x| > 1$. 

\begin{figure}
\centering
\includegraphics[width=0.5\textwidth]{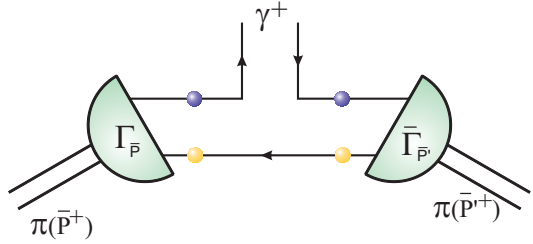}
\caption{Schematic representation of the expression of the pion quark
  GPD in terms of Bethe-Salpeter vertex functions, Eq.~(\ref{eq:Hq_vertex}).}
\label{Fig:hq_bs}
\end{figure}

In a Bethe-Salpeter quark model, the bilocal current matrix element
from Eq.~(\ref{eq:defgpd}) can be written in terms of the Bethe-Salpeter amplitudes $\chi_{\bar{P}}$:  
\begin{equation}
\begin{split}
  \langle\bar{P}^{\prime}|\bar{\psi}&(-\frac{z}{2})\gamma^{+}\psi(\frac{z}{2})|\bar{P}\rangle\\ 
  &=
  \int\mathrm{d}^4x_2'~i\mathrm{Tr}\left\{\bar{\chi}_{\bar{P}^{\prime}}(x_2',\frac{z}{2})
  \left((i\feyndag{\partial}_{x_2'}-m_1)\chi_{\bar{P}}(x_2',-\frac{z}{2})
  \right)\gamma^{+}\right\}\\ 
  &+ 
  \int\mathrm{d}^4y_1'~i\mathrm{Tr}\left\{\left((i\feyndag{\partial}_{y_1'}-m_2)
  \bar{\chi}_{\bar{P}^{\prime}}(-\frac{z}{2},y_1')\right)\gamma^{+}
  \chi_{\bar{P}}(\frac{z}{2},y_1')\right\}\,.  
\end{split}
\label{eq:matrixelement_bs}
\end{equation}
The first (second) term on the right-hand side of this equation refers
to the coupling of the bilocal current to the antiquark (quark). Both
terms can be treated in exactly the same manner. In the remainder of
this letter, we focus on the analysis of the second term of
Eq.~(\ref{eq:matrixelement_bs}). Inserting the quark term in
Eq.~(\ref{eq:defgpd}) and making use of the definition of the quark
and antiquark propagators, 
\begin{equation}
(i\feyndag{\partial}_x-m)S^F(x) = i\delta^{(4)}(x-x')\,, 
\label{eq:prop_diffeq}
\end{equation}
we can rewrite the quark GPD as an integral in momentum space:
\begin{equation}
  \begin{split}
    H_{\pi}^{q}&(x,\xi,t) = 
    -\frac{1}{2}\int\frac{\mathrm{d}^4p}{(2\pi)^4} \delta
    \left(\frac{2x+\xi-1}{2(1+\xi)} \bar{P}^+-{p}^+ \right)\\   
    &\times\mathrm{Tr}\left\{\bar{\Gamma}_{\bar{P}^{\prime}}(p + \frac{\Delta}{2})
    S^F_{1}(\frac{\bar{P'}}{2} + p + \frac{\Delta}{2})\gamma^+S^F_{1}
    (\frac{\bar{P}}{2}+p)\Gamma_{\bar{P}}(p) S^F_{2}(-\frac{\bar{P}}{2} +
    p)\right\} \,.   
    \label{eq:Hq_vertex}
  \end{split}
\end{equation}
On the left-hand side of Eq.~(\ref{eq:Hq_vertex}), the superscript
$q$ indicates that only the coupling to the quark is considered. The
vertex functions 
\begin{equation}
\Gamma_{\bar{P}} = (S_1^F)^{-1}\chi_{\bar{P}}(S_2^F)^{-1}
\end{equation}
have been introduced on the
right-hand side of the equation, with subscript $1$ ($2$) referring to
the quark (antiquark). The pictorial representation of
Eq.~(\ref{eq:Hq_vertex}) is sketched in Fig.~\ref{Fig:hq_bs}. In some
Bethe-Salpeter models, see e.g. Ref. \cite{koll00}, the quark and antiquark
propagators are assumed to have the form of a free propagator of
pointlike fermions, with a constituent mass $m$: 
\begin{equation}
S_{1,2}^F(p) = i\frac{\feyndag{p} + m_{1,2}}{p^2 -
  m_{1,2}^2 + i\epsilon}\,.
\label{eq:freeprop}
\end{equation}
As it appears, the denominators of the three propagator terms in
Eq.~(\ref{eq:Hq_vertex}) suggest immediately the correct support
region for the GPDs. Since the poles of the propagators are Lorentz
invariants, any convenient frame can be chosen to prove this
statement. The GPD in Eq.~(\ref{eq:defgpd}) was defined for a
coordinate system with $\vec{\tilde{P}}$ along the $z$-axis
\cite{ji:first}. The Breit frame is such a coordinate system, with
$\bar{P} = (\bar{M},-\frac{\vec{\Delta}}{2})$ and $\bar{P}^{\prime} =  
(\bar{M},\frac{\vec{\Delta}}{2})$. Here, $\bar{M}^2 = M^2 -
\frac{t}{4}$ and $\vec{\Delta} = (\vec{\Delta}_{\perp}, -2\xi\bar{M})$
with $\xi$ as introduced before. 
We emphasize that these expressions uniquely apply to the
calculation of the \emph{elastic} GPD, $M' = M$.   

Making use of the $\delta$-function to perform the integration over
$p^+$, it is now possible to write the propagator 
denominators as: 
\begin{eqnarray*}
  \big((\frac{\bar{P}}{2}+ p + \Delta)^2 - m_1^2 + i\epsilon\big)&=&
  \mathcal{A} + \sqrt{2}\bar{M}(x-\xi)p^- + i\epsilon\,,\\
  \big((\frac{\bar{P}}{2} + p)^2 - m_1^2 + i\epsilon\big) &=&
  \mathcal{B} + \sqrt{2}\bar{M}(x+\xi)p^- + i\epsilon\,,\\
  \big((-\frac{\bar{P}}{2} + p)^2 - m_2^2 + i\epsilon\big)&=& 
  \mathcal{C} + \sqrt{2}\bar{M}(x-1)p^- + i\epsilon\,,
\end{eqnarray*}
where $\mathcal{A}$, $\mathcal{B}$ and $\mathcal{C}$ do not depend on
$p^-$. From these expressions, it is clear that the propagators 
have poles in the complex $p^-$-plane. The position of these poles
with respect to the real $p^-$-axis (upper or lower half-plane)
depends on the kinematics, more specifically on the sign of $(x-\xi)$, 
$(x+\xi)$ and $(x-1)$. For $x\notin[-\xi,1]$, the propagator poles lie
on the same side of the real axis. Furthermore, the propagators are
analytic in all other points of the complex plane. Assuming that the
vertex function $\Gamma$ does not contain poles in the complex
$p^-$-plane, and making use of Cauchy's theorem, one can conclude that
the momentum dependence of the propagators ensures the correct support
for the GPDs.

The above considerations are clearly valid if the vertex function
$\Gamma$ and its adjoint do not depend on $p^-$ and are therefore free
from poles when analytically continued in the complex $p^-$-plane. The 
calculation of the pion GPD with constant vertex functions has been
performed by S.~Noguera et al. in the NJL-model \cite{noguera:bsapp}
with a Pauli-Villars regularization procedure. The calculation of GPDs
with $p^{-}$-independent vertex functions was performed by H.-M.~Choi
et al. \cite{choi1,choi2}, as well as by B.C.~Tiburzi and G.A.~Miller
\cite{tibmil1,tibmil2,tibmil3}. These authors made use of a reduction
of Bethe-Salpeter amplitudes to light-cone wave functions by
projection on the light-cone. In Refs. \cite{tibmil1,tibmil2,tibmil3}, the
scalar Wick-Cutkosky model was adopted and successfully applied to the
calculation of scalar meson GPDs. In Refs. \cite{choi1,choi2}, the
light-front Bethe-Salpeter vertex functions were replaced by wave
functions obtained in a light-front constituent quark model. All of
these calculations yielded the correct support, in agreement with the
present analysis. 

In instant-form models, however, the vertex functions can depend on
$p^-$, which makes them have a different pole structure. In this case,
the correct support can no longer be guaranteed. To prove this, we
start with Liouville's theorem, which states that \emph{the only
  bounded entire functions are the constant functions}. As a result, 
a vertex function which depends on the momentum variable $p^-$ will
not be bounded for all points in the complex $p^-$-plane. Moreover,
the radial part of a vertexfunction for a ground-state particle is a
real-valued function on the real $p^-$ axis, up to a constant phase
factor. This is an important constraint. Indeed, for a holomorphic
function $f$ whose restriction to the real numbers is real-valued it
can be proven that $f(p^{-*}) = f^*(p^-)$. Here, $^*$ stands for
complex conjugation. In other words: if $f$ has a singularity in the
upper half-plane, it also has one in the lower half-plane, and vice
versa. Combining these two statements, we find that a dynamical,
momentum dependent vertex function will either have complex conjugated
poles, or a singularity on both complex  half-circles at $x\pm i
\infty$. In the latter case, Cauchy's theorem is no longer applicable,
while in the former case, there will always be at least one pole in
each half-plane. Consequently, it is not a priori clear whether the
GPD will vanish outside the interval $x\in[-1,1]$. This means that the
correct support can no longer be guaranteed. Similar arguments apply
to relativistic quark-diquark models of the nucleon with pointlike
constituents.

\begin{figure}
\centering
\includegraphics[width=0.5\textwidth]{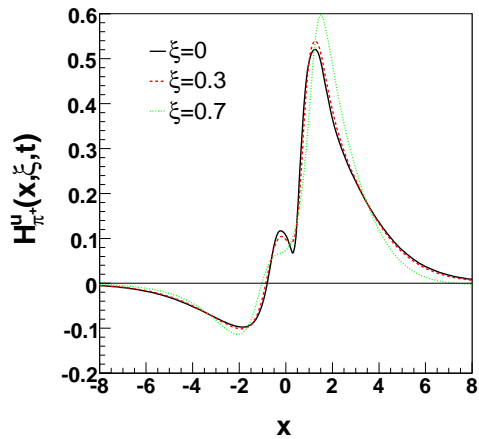}
\caption{The pion GPD $H_{\pi^+}^{u}$ as a function of $x$, calculated
  in the Bonn constituent quark model at $t = -0.1$ GeV$^2$. Results
  are shown for $\xi$ are $\xi = 0$
  (solid line), $\xi = 0.3$ (dashed line) and $\xi = 0.7$ (dotted
  line).}
\label{Fig:plot}
\end{figure}

We stress that this does not mean that there will
necessarily be a support problem in Bethe-Salpeter constituent quark
models with pointlike constituents containing dynamics - it merely
shows that a support problem can \emph{no longer be excluded}. The
above considerations indicate that guaranteeing the correct
support puts non-trivial constraints on the analytic properties of the
Bethe-Salpeter vertex functions. Whether these are fulfilled in
particular models has to be checked in each case individually.  
As an example, we show in Fig. \ref{Fig:plot} the
pion GPD calculated in the framework of the covariant quark model of
the Bonn group \cite{koll00}. 

In the Bonn model, the interaction kernels are 
instantaneous, \emph{i.e.} independent from relative energy variables
in the meson rest frame. More specifically, the employed interquark
interactions are the linearly rising confinement interaction and the
't Hooft instanton induced interaction \cite{tHooft}. The confinement
interaction has an appropriate Dirac structure to optimize the
splitting between spin-orbit partners. The 't Hooft 
instanton induced interaction acts as a residual interaction and
accounts for the mass splittings in the pseudoscalar and scalar
sectors \cite{metsch96}. Together with the assumption of free fermion
propagators with an effective mass (Eq. (\ref{eq:freeprop})), the
instantaneous approximation allows to reduce the full (but unsolvable)
Bethe-Salpeter equation to a solvable Salpeter equation
\cite{resag95}. The resulting vertex function is independent of the
relative energy variable, but does depend on the relative
minus-momentum. The Bonn model has only seven 
parameters which are fitted to the static properties of the meson
spectrum. In Refs. \cite{koll00,resag95,muenz94,muenzthesis},
the model is described in detail.

Figure \ref{Fig:plot} shows the result of a calculation of the pion
up-quark GPD $H_{\pi^+}^{u}$ as a function of $x$ at $t = -0.1$ GeV$^2$
and three different values for $\xi$ ($\xi=0$, $\xi=0.3$ and
$\xi=0.7$). 
The first moment of the GPD, 
\begin{equation}
\int_{-\infty}^{\infty} dxH_{\pi^+}(x,\xi,t)
\end{equation}
yields the electromagnetic form factor, while the isospin symmetry relation
\cite{diehl:review},
\begin{equation}
  H_{\pi^+}^{u}(x,\xi,t)=-H_{\pi^+}^{d}(-x,\xi,t)\,,
\label{Eq:isospinsymm}
\end{equation}
is exactly fulfilled. Here, $H_{\pi^+}^{d}$ is calculated from the
first term on the right-hand side of Eq.~(\ref{eq:matrixelement_bs}),
which describes the coupling of the bilocal current to the
antiquark. As a measure of the support problem,
the support parameter $\phi$ can be introduced:  
\begin{equation}
\phi = \frac{\int_{-\xi}^1 |H_{\pi^+}^{u}(x,\xi,t)|dx}{\int_{-\infty}^{\infty}
  |H_{\pi^+}^{u}(x,\xi,t)|dx}\,.
\end{equation}
When the appropriate kinematical regions are resolved (see
Fig. \ref{Fig:regions}), this fraction equals one. In contrast, $\phi < 1$
implies that the GPD is non-zero outside the supported region, which
is $[-\xi,1]$ for the quark term. The smaller the fraction, the worse
the support. In the case of Fig. \ref{Fig:plot}, this fraction equals
$\phi = 0.13$ for $\xi=0$ and $\xi=0.3$. For $\xi=0.7$, one has $\phi =
0.12$. From these numbers, it is clear that the correct support is
violated in the Bonn constituent quark model.

\section*{Acknowledgments}
The authors wish to thank B. Metsch for fruitful
discussions and for carefully reading the manuscript. Enlightening
discussions with S. Scopetta and D. Van Neck are gratefully
acknowledged. This work was supported by the Research Foundation -
Flanders (FWO).  

\bibliographystyle{physlettb}
\bibliography{/home/annelies/Work/Articles/MyArticles/MesonGPDs/Letter/Ingestuurd/v3/references.bib}

\newpage

\end{document}